\documentclass[aps,prl,twocolumn,amsmat,amssymb,amsfonts,superscriptaddress]{revtex4-1}
\usepackage{amsmath}
\usepackage{times}
\usepackage{hyperref}

\usepackage{color}
\usepackage{graphics}
\usepackage{graphicx}
\usepackage[percent]{overpic}
\usepackage{wasysym}
\begin{document}

\title{Non-local String Order Parameter in the $S=1/2$  Kitaev-Heisenberg Ladder}
\author{Andrei Catuneanu}
\affiliation{Department of Physics, University of Toronto, Ontario M5S 1A7, Canada}
\author{Erik S. S{\o}rensen}
\email{sorensen@mcmaster.ca}
\affiliation{Department of Physics, McMaster University, Hamilton, Ontario L8S 4M1, Canada}
\author{Hae-Young Kee}
\email{hykee@physics.utoronto.ca}
\affiliation{Department of Physics, University of Toronto, Ontario M5S 1A7, Canada}
\affiliation{Canadian Institute for Advanced Research, CIFAR Program in Quantum Materials, Toronto, ON M5G 1M1, Canada}
\date{\today}

\begin{abstract}
We study the spin-$\frac{1}{2}$ Kitaev-Heisenberg (KJ) model in a two-leg ladder. 
Without a Heisenberg interaction, the Kitaev phase in the ladder model has Majorana fermions with local Z$_2$ gauge fields,
and is usually described as a disordered phase without any order parameter. 
Here we prove the existence of a non-local string order parameter (SOP) in the Kitaev phase which survives with a finite Heisenberg interaction.
The SOP is obtained by relating the Kitaev ladder, through a non-local unitary transformation, to a one-dimensional $XY$ chain with an Ising coupling to a dangling spin at every site. 
This differentiates the Kitaev phases from other nearby phases including a rung singlet.
Two phases with non-zero SOP corresponding to ferromagnetic and antiferromagnetic Kitaev interactions are identified. 
The full phase diagram of the KJ ladder is determined using exact diagonalization and density matrix renormalization group methods,
which shows a striking similarity to the KJ model on a two-dimensional honeycomb lattice. 
\end{abstract}
\maketitle

{\it Introduction} -- Topological quantum phase transitions (QPT) at zero temperature do not involve any local order parameter. 
Due to the absence of a local order parameter, conventional Landau theory fails and characterizations of such transitions has become one of the challenging tasks in modern condensed matter physics.
In one-dimensional (1D) systems, a topological QPT may be accompanied by a non-local string order parameter (SOP).
The best example is the spin $S=1$ Haldane phase~\cite{Haldane1983a,Haldane1983b}.
A feature of the Haldane phase is the breaking of a hidden $Z_2 \times Z_2$ symmetry revealed by a SOP defined through a non-local unitary transformation~\cite{denNijs1989,Kennedy1992a,Kennedy1992b,Oshikawa1992}.
However, identifying a relevant SOP in $S=\frac{1}{2}$ ladder systems is a non-trivial task, particularly for highly frustrated spin interactions,
although heuristic extensions of the $S=1$ SOP to $S=\frac{1}{2}$ ladders have been discussed~\cite{White1996,Nishiyama1996,Kim2000,Fath2001,Anfuso2007,SM}.

\begin{figure}[!t]
\includegraphics[width=0.9\linewidth,clip]{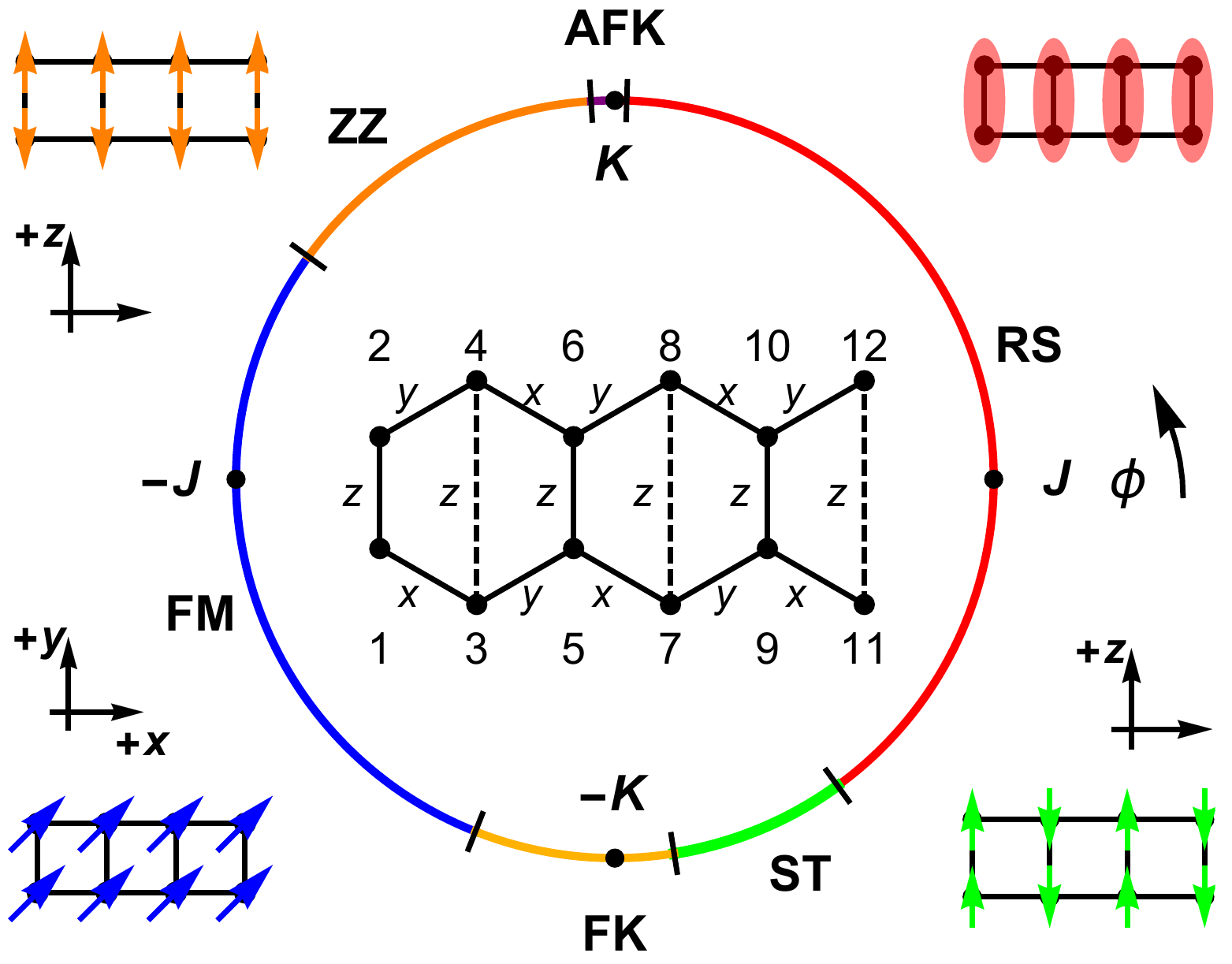}
\caption{
\label{fig:phasediagram}
(Color online) 
  Phase diagram of the KJ model as a function of $\phi$ with numerically determined phase boundaries 
labeled by black bars. 
Six phases are identified. The rung singlet (RS) is a singlet state, and the ZZ and ST phases have a ferromagnetic Ising ordering only along
a leg and rung, respectively shown by colored spins
with accompanying quantization axes. FM has a ferromagnetic long range order.
AFK and FK are the Kitaev phases. The phase transition boundaries are similar to those of the two-dimensional (2D) KJ model.
The two-leg ladder with bond definitions for the Kitaev term in Eq. \ref{eq:hkj} is depicted underneath with bond-dependent interactions denoted by
x, y, and z.
}
\end{figure}

We study the $S=\frac{1}{2}$ Kitaev-Heisenberg (KJ) model in a two-leg ladder.
The Kitaev model is described by bond-dependent interactions between nearest neighbors on a two-dimensional (2D) honeycomb lattice~\cite{Kitaev2006}.
Taking two rows of the honeycomb lattice
and connecting the dangling bonds (dashed lines in Fig. ~\ref{fig:phasediagram}) generates a two-leg ladder with bond-dependent
interactions.
A previous study\cite{Feng2007} showed that
this simplified ladder captures the exact phase transition boundaries between the two gapped ($A_x$, $A_y$) and a gapless ($B$) spin liquid occurring in the 2D honeycomb limit~\cite{Kitaev2006},
despite the fact that all the equivalent phases are gapped in the ladder geometry. 
The naming convention of phases follows Ref.~\cite{Kitaev2006}.
One of the gapped phases, which becomes the gapless ($B$) Kitaev spin liquid in the 2D limit, 
was characterized as a disordered phase without any order parameter~\cite{Feng2007,McCoyPRA1971}.

Here we prove the existence of a long-range {\it non-local} SOP in this phase which survives with a finite Heisenberg interaction.
The full phase diagram of the Kitaev-Heisenberg (KJ) model on the ladder is determined using the exact diagonalization (ED) and density matrix renormalization group (DMRG)
techniques. 
A striking similarity to the 2D phase diagram on the honeycomb lattice\cite{ChaloupkaPRL2013} is found, despite the different geometries.
The SOP differentiates the Kitaev phase from a rung singlet, and 
other phases corresponding to the zig-zag, stripy, and ferromagnetic phases found in the 2D limit are also captured in the ladder.

{\it Phase diagram} --
The KJ Hamiltonian defined on a two-leg ladder is given by
\begin{equation}
H = K \sum_{\gamma \in \langle i, j\rangle} S^\gamma_i S^\gamma_j + J \sum_{\langle i, j\rangle} {\bf S}_i \cdot {\bf S}_j,\label{eq:hkj}
\end{equation}
where $S=1/2$, $\langle i,j\rangle$ are site indices defined on nearest-neighbor bonds, and $\gamma = x,y$ or $z$ depending on bond type as shown in Fig. \ref{fig:phasediagram}.
The first term is the bond-dependent Kitaev interaction while the second is the isotropic Heisenberg interaction. 
We parameterize the spin exchanges by $K=\sin\phi$ and $J=\cos\phi$ where $\phi \in [0,2\pi)$.
 When $\phi = \frac{\pi}{2}$ or $\frac{3\pi}{2}$, the Hamiltonian reduces to the Kitaev ladder studied in Ref. \cite{Feng2007}. 
   
We first determine the entire phase diagram of the Hamiltonian as a function of $\phi$.
We have numerically diagonalized the KJ model using ED on a $N=24$ site ladder using periodic boundary conditions (PBC),
and DMRG.
Phase boundaries were determined by identifying singular features of the second derivative of the ground state energy per site, $\chi_E=-\partial^2_\phi e_0$, 
the presence of a gap and the presence of a non-zero SOP.
Six different phases were identified as shown in Fig. \ref{fig:phasediagram}:
(a) a rung singlet (RS)~\cite{Dagotto1992,Barnes1993,White1994}  phase with a gap, 
(b) an easy-plane ferromagnetic (FM) phase,  
(c) a gapped phase with opposing long-range FM Ising order on each leg (ZZ), 
(d) a gapped phase with alternating long-range FM Ising order on the rungs (ST), 
and (e) antiferromagnetic Kitaev (AFK) and (f) FM Kitaev (FK) phases.  

The AF and FM Heisenberg limits at $\phi = 0$ and $\pi$ are located in the RS and FM phases (black
dots in Fig. \ref{fig:phasediagram}). 
In the thermodynamic limit, the ZZ, ST, AFK and FK phases all have two-fold degenerate ground-state while the RS phase has a unique ground-state.
We estimate the other transitions as: RS-AFK: $\phi\simeq 0.487\pi$, AFK-ZZ: $\phi\simeq 0.53\pi$, ZZ-FM:$\simeq 0.81\pi$, FM-FK: $\phi\simeq 1.377\pi$, FK-ST: $\phi\simeq 1.563\pi$ and ST-RS:$\phi\simeq 1.71$.
We note that despite the
manifestly 1D nature of the ladder geometry, the phase diagram is strikingly
similar to that of the 2D honeycomb phase diagram for the Kitaev-Heisenberg model that has been previously
studied~\cite{Chaloupka2010,Gotfryd2017,Agrapidis2018} and we have
chosen the naming of the ZZ and ST magnetically ordered phases to correspond.
The only qualitative difference though is that at the AFK to ZZ transition no feature is observed~\cite{SM} in $\chi_E$ nor in the fidelity susceptibility in the ladder, implying that the transition is likely high order
with $2/\nu-d < 0$~\cite{Albuquerque2010} or $\nu>2$, which is different from the 2D honeycomb.
A level-crossing between
the two ground-states, split by finite-size effects, is however observed at
$\phi\simeq 0.515\pi$~\cite{SM}. 
A further discussion on this transition is presented later.

Let us first focus on the nature of Kitaev phases near $\phi=\frac{\pi}{2}$ and $\frac{3\pi}{2}$.
The ground states of the ladder at the Kitaev points, with $\pm K$, have been described as disordered without a SOP~\cite{Feng2007,McCoyPRA1971}.
Here we demonstrate the existence of a long-range SOP in both AFK and FK phases with and without Heisenberg interaction.
We shall do this by explicitly establishing a {\it non-local} unitary
transformation $V$ that maps $H$
to another Hamiltonian with a non-zero {\it local} order parameter.
Applying the inverse transformation to this local order parameter then yields the (hidden) SOP in the original model.
As we shall demonstrate, this SOP differentiates the Kitaev phases from neighboring phases.

{\it The Unitary operator $V$} -- 
Previous
studies~\cite{Kennedy1992a,Kennedy1992b,Oshikawa1992} have exclusively
considered $S=1$ models for the technical reason that the integer spin identity
$\exp(2 i\pi S_j^x S_k^z)=I$ is not satisfied for $S=\frac{1}{2}$. However, here we
show that even without this identity one can still define a suitable unitary
operator for the Kitaev ladder.  In order to define such a unitary operator,
we group the $S=\frac{1}{2}$ spins in pairs. While it is possible to group any two
spins into a pair, we  make the simple choice to group spins on the rungs
of the ladder. 
Following the numbering convention of the lattice sites shown in Fig.~\ref{fig:phasediagram}
we then define the following non-local unitary operator for a
$N$-site ladder with open boundary conditions (OBC):
\begin{equation}
  V = \prod_{\substack{j+1< k \\ j\ \mathrm{odd},\  k\ \mathrm{odd} \\ j=1,\ldots N-3 \\ k=3,\ldots N-1}} U(j,k),
\end{equation}
where the individual $U(j,k)$ given as follows:
$  U(j,k)  = e^{i\pi(S^y_j+S^y_{j+1})(S^x_k+S^x_{k+1})}$.
Clearly, all $U(j,k)$ are unitary and therefore also $V$; and, as mentioned above, $j, j+1$ and $k, k+1$ group the $S=
\frac{1}{2}$ spins on a rung. 
We note that,
$  \left[U(j,k),U(l,m)\right]  = 0 \ \ \forall \ j,k,l,m$
which allows us to rearrange terms in a convenient manner. 

\begin{figure}[h]
\includegraphics[width=0.9\linewidth]{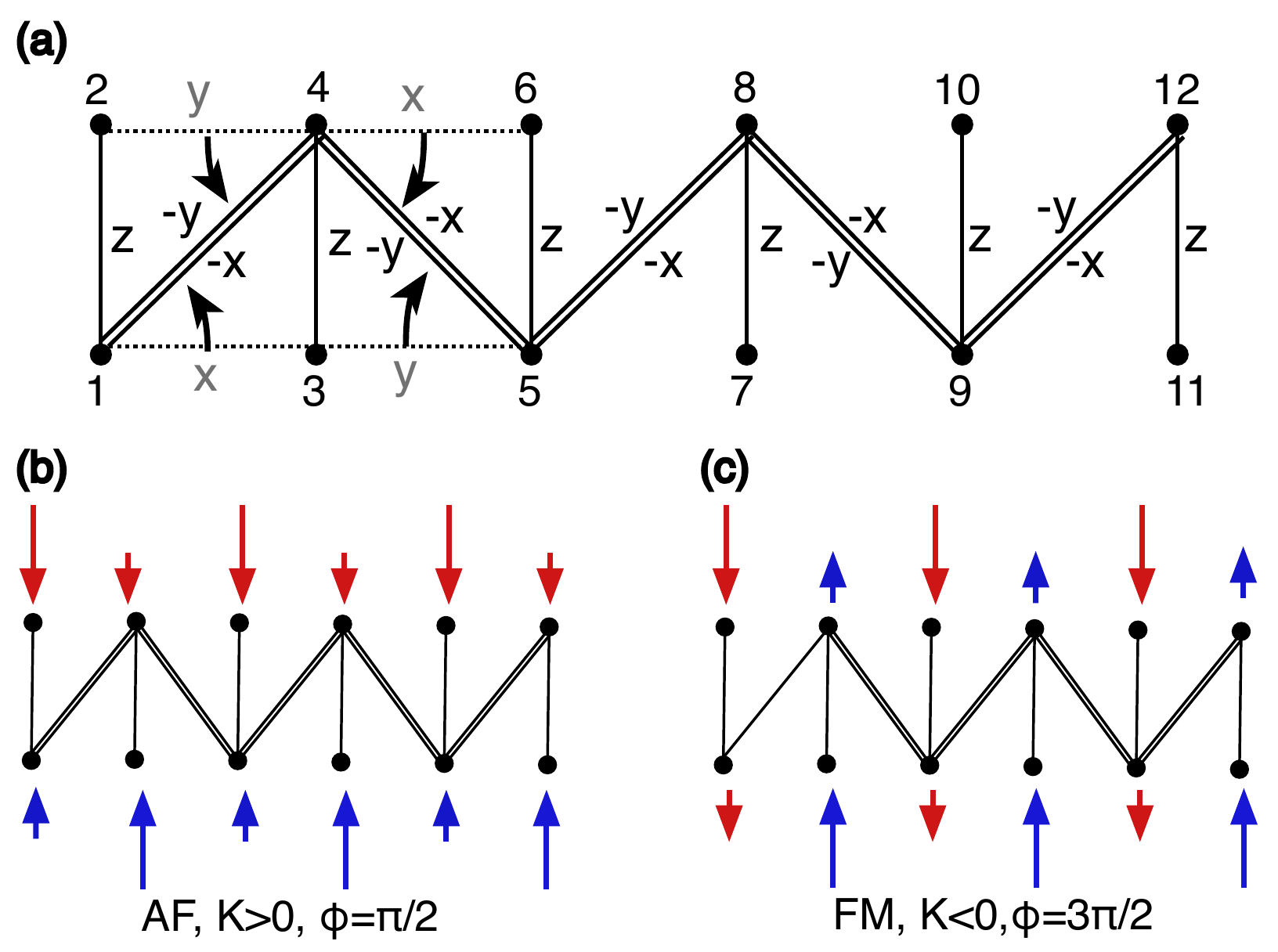}
\caption{
  \label{fig:DanglingZ} 
   (Color online)
(a) The movement of bonds by the unitary operator $V$ is indicated in the first part of the chain.
Schematic view of (b) the ordering at the AF Kitaev point, $\phi=\frac{\pi}{2}$, and
(c) the FM Kitaev point, $\phi=\frac{3\pi}{2}$ for the $H_{d-Z}$. 
 In each case one of 2 degenerate ground states is shown, with the other one obtained by interchanging
up and down spins on the dangling spin sites.
}
\end{figure}

Under the unitary transformation $V$ the individual spin operators
$S^\alpha_i$ will transform in a way that depends on $i$ as shown in the supplementary materials (SM).
Note that $V$ effectively {\it moves} the bond and changes the sign of the interaction as sketched in the Fig.~\ref{fig:DanglingZ}(a).
Using the numbering of the sites as in Fig.~\ref{fig:phasediagram}, we find that the interactions around a plaquette transform under $V$ as follows:
\begin{eqnarray}
  VS_{4n+1}^xS_{4n+3}^xV^{-1} &=& -S_{4n+1}^x S_{4n+4}^x \nonumber\\
  VS_{4n+2}^xS_{4n+4}^xV^{-1} &=& -S_{4n+2}^x S_{4n+3}^x,
\end{eqnarray}
with $n =0,1,2,...$.
For the $y-y$ interaction, it changes to
\begin{eqnarray}  
  VS_{4n+1}^yS_{4n+3}^yV^{-1} &=& -S_{4n+2}^y S_{4n+3}^y \nonumber\\
  VS_{4n+2}^yS_{4n+4}^yV^{-1} &=& -S_{4n+1}^y S_{4n+4}^y,
\end{eqnarray}
On the other hand, the $z-z$ interaction on the rungs, i.e., $S_i^z S_{i+1}^z$ for $i=1,3,5...$ is unchanged.
Other transformations of interaction terms are given in the SM. 

With these transformations we see that the original Kitaev Hamiltonian in Eq.~(\ref{eq:hkj}) (with $J=0$ and OBC) is transformed
as $VHV^{-1}=H_{d-Z}$ with:
\begin{eqnarray}
  H_{d-Z} = & &K\sum_{n=0}\tilde S_{2n+1}^z\tilde S_{2n+2}^z\nonumber\\
  &&-K\sum_{n,\alpha=x,y}\left(
  \tilde S_{4n+1}^\alpha\tilde S_{4n+4}^\alpha
  +\tilde S_{4n+4}^\alpha\tilde S_{4n+5}^\alpha\right),
  \label{eq:HdZ}
\end{eqnarray}
where $\tilde S$ denotes the spins in $H_{d-Z}$.
The transformed Hamiltonian is essentially an $XY$ chain with Ising coupling to a dangling spin at every site,
which we therefore name the ``dangling-Z'' model.

$H_{d-Z}$ has several interesting properties, most importantly, all the ``dangling" $\tilde S^z$ commute with $H_{d-Z}$. With our numbering:
\begin{equation}
  \left[\tilde S^z_{4n+2},H_{d-Z}\right]=0, \ \ \left[\tilde S^z_{4n+3},H_{d-Z}\right]=0.\ \ n=0,1,2
\end{equation}
Hence,  each eigenstate of $H_{d-Z}$ will be part of a $2^{N/2}$ manifold of
states generated by the different configurations of the free $\tilde S^z$ spins. All
$2^{N/2}$ states are two-fold degenerate, corresponding to flipping all the dangling spins. Sketches
of one of the two
ground-states at the Kitaev points with the dangling-Z spins fully polarized are
shown in Fig.~\ref{fig:DanglingZ}(b) and (c).
Following Kitaev's idea, one can represent $H_{d-Z}$ in terms of Majorana operators. 
  $H_{d-Z}$ can be mapped to free Majorana fermions along the deformed zig-zag chain which couple to a Z$_2$ flux at dangling sites via the $\tilde{S}^z\tilde{S}^z$ interaction as shown in the SM. Thus the $2^{N/2}$ manifold of states can be understood in terms of Z$_2$ flux degrees of freedom. 

\begin{figure}[h]
        \includegraphics[width=0.95\linewidth]{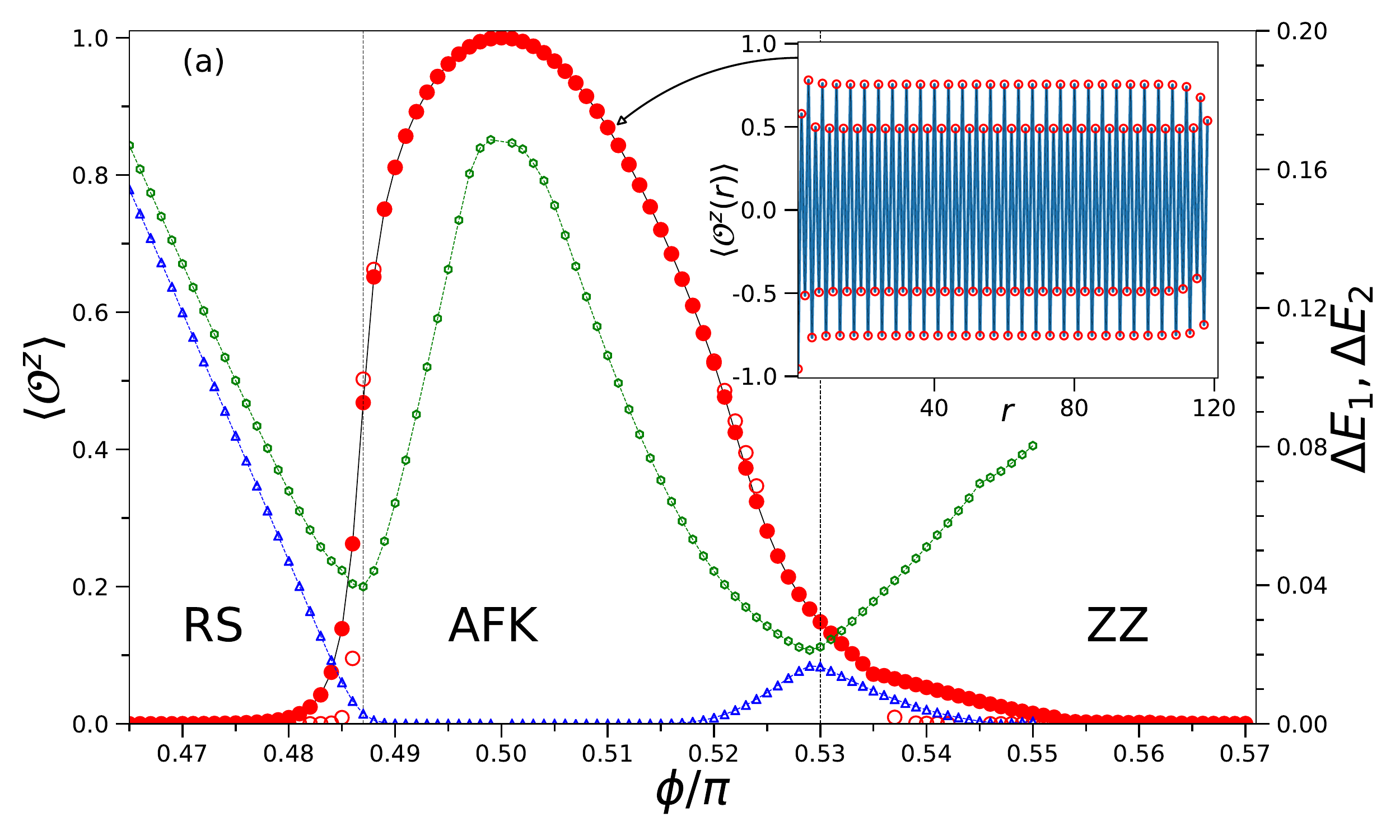}
          \includegraphics[width=0.95\linewidth]{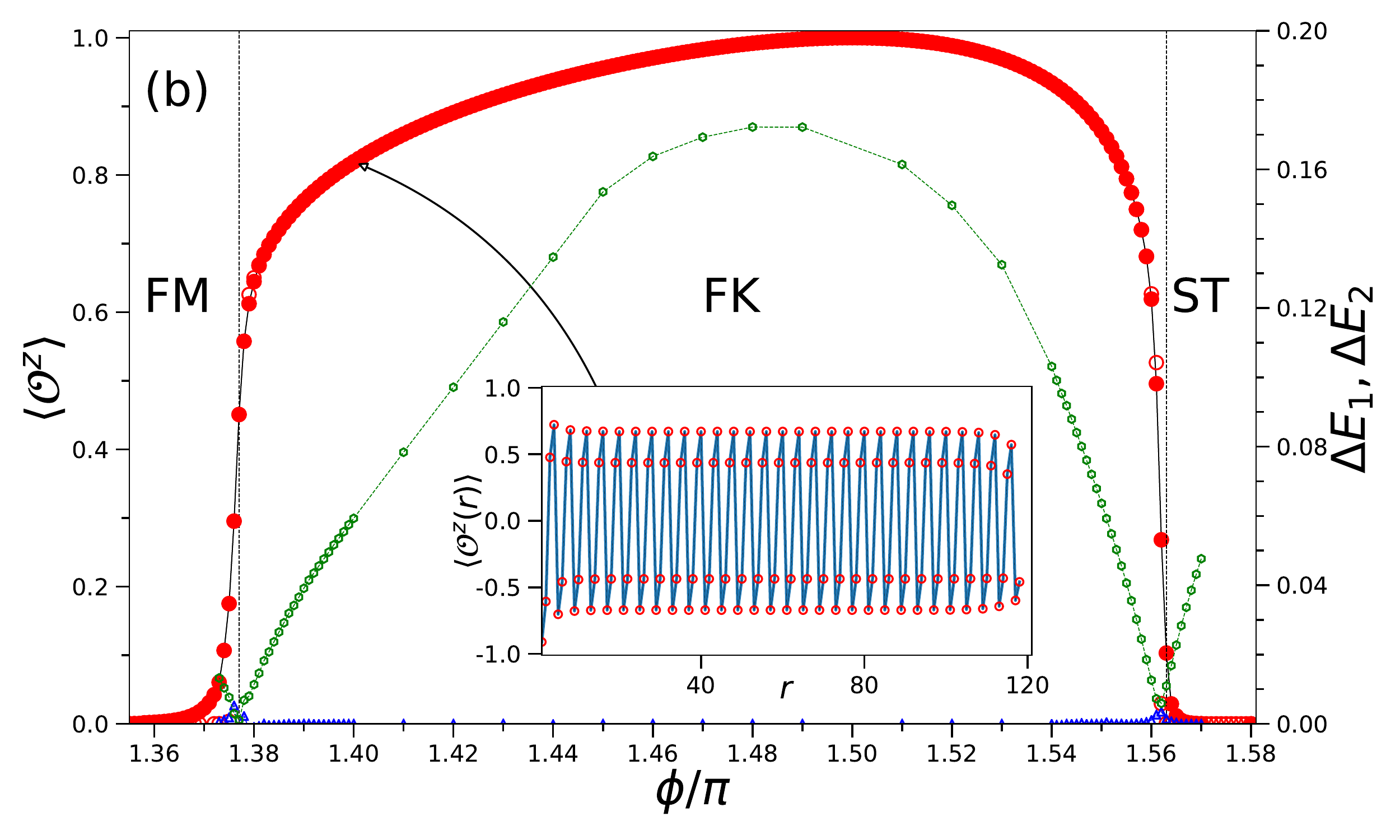}
        \caption{\label{fig:AFKLOzString} (Color online) $\langle {\cal O}^z\rangle $ determined from DMRG calculations on a ladder with $N=120/400$ sites 
         (solid/open red circles) and OBC with a typical truncation error of $10^{-11}$.
        (a) near the AFK (Inset for $\langle {\cal O}^z(r)\rangle$ vs $r$ at $\phi=0.51\pi$),
        and (b) near the FK (Inset for  $\langle {\cal O}^z(r)\rangle $ vs. $r$ at $\phi=1.40\pi$).
        The energy gap to the first excited state, $\Delta E_1$  and
         the second excited state $\Delta E_2$ are shown by blue triangles and green circles, respectively. Both gaps are determined by
         DMRG calculations with $N=60$ and PBC with a typical truncation error of $10^{-9}(10^{-6})$ on the ground (excited) state. 
        }
\end{figure}
{\it String Order Parameter} --
With the dangling spin integrals of motion it is clear that $H_{d-Z}$ can have long-range order in the sense
that $\lim_{r\to\infty} \langle \tilde S_i^\alpha\tilde S_{i+r}^\alpha\rangle \neq 0$, $\alpha=x,y,z$.
We can then define string correlation functions in the original $H$, Eq.~(\ref{eq:hkj}), that are ordinary correlation functions in the 
transformed Hamiltonian, $H_{d-Z}$.
We define a $z-$string correlation function starting from the leftmost dangling site 2:
\begin{eqnarray}
  \langle{\cal O}^z(r)\rangle &=& 4 \langle \tilde S_2^z\tilde S_{2+r}^z\rangle =  (-1)^{\left \lfloor{(r+1)/2}\right \rfloor }\nonumber\\
  &\times&\left\{ \begin{array}{rl}
  \langle\sigma_1^y\sigma_2^x \left( \prod_{k=3}^{r+1}\sigma_k^z \right)\sigma_{2+r}^x\sigma_{3+r}^y\rangle &\mbox{ $r$ odd} \\
    \\
  \langle\sigma_1^y\sigma_2^x \left( \prod_{k=3}^{r+1}\sigma_k^z \right)\sigma_{2+r}^y\sigma_{3+r}^x\rangle &\mbox{ $r$ even},
       \end{array} \right.\label{eq:ozr}
\end{eqnarray}
where $\mathbf{\sigma}_i $ are the Pauli matrices in the original $H$.
Note that ${\cal O}^z(r)$ contains a combination of $x,y,z$ Pauli matrices.
With this definition, long-range order in $\langle \tilde S_i^z\tilde S_{i+r}^z\rangle$ results in long-range order in ${\cal O}^z(r)$.
Similarly, an $x$-string operator that starts in the leftmost site 1 is found:
\begin{eqnarray}
  \langle{\cal O}^x(r)\rangle &=& 4 \langle \tilde S_1^x\tilde S_{1+r}^x\rangle =  (-1)^{\left \lfloor{(r+1)/2}\right \rfloor }\nonumber\\
  &\times&\left\{ \begin{array}{rl}
  \langle\sigma_1^x \left( \prod_{k=3}^{r-1}\sigma_k^x \right)\sigma_{r}^x\rangle &\mbox{ $r$ odd} \\
    \\
  \langle\sigma_1^x \left( \prod_{k=3}^{r}\sigma_k^x \right)\sigma_{r+2}^x\rangle &\mbox{ $r$ even} 
       \end{array} \right.\label{eq:oxr}
\end{eqnarray}
with a similar expression for the $y$-string operator. Note that, in this case the string of $\sigma^x$'s is not consecutive. 

It is clear that ${\cal O}^z(r)$ is long-ranged at the Kitaev points, $\phi=\pi/2$ and $3\pi/2$, because of the dangling $\tilde S^z$ local integrals of motion.
However, in the presence of a non-zero Heisenberg term $J$, new terms arise in $VHV^{-1}$ (see supplemental material) which is no longer simply equal to Eq.~(\ref{eq:HdZ}).
It is therefore not at all obvious that ${\cal O}^z(r)$ will show long-range order.
 To understand the hidden order near the Kitaev points with $J\neq 0$, 
we define an associated string order parameter as follows:
\begin{equation}
  {\cal O}^z = \sqrt{|{\cal O}^z(3L/4)|_{max}}
\end{equation}
where $|{\cal O}^z(3L/4)|_{max}$ refers to the maximal value ${\cal O}^z(r)$ takes in the neighborhood of $r=3L/4$. This definition avoids effects from
the open boundary at $r=L$.
With this definition of an SOP, we can now map out the phase
around $\phi=\pi/2$ and $\phi=3\pi/2$ where the ``hidden" order associated with
the z-string correlation functions ${\cal O}^{z}(r)$ is present.  
On the other hand, the x- and
y-string correlation functions ${\cal O}^{x/y}(r)$ show exponentially decaying
behavior at the Kitaev points, ${\cal O}^{x/y}(r)=ae^{-r/\xi}$ with $\xi\sim 4$~\cite{SM}, and ${\cal O}^{x/y}$ is not
long-ranged in the FK and AFK phases although it is trivially long-range in the
FM phase where instead ${\cal O}^{z}$ is zero.

\begin{figure}[h]
\includegraphics[width=1.05\linewidth]{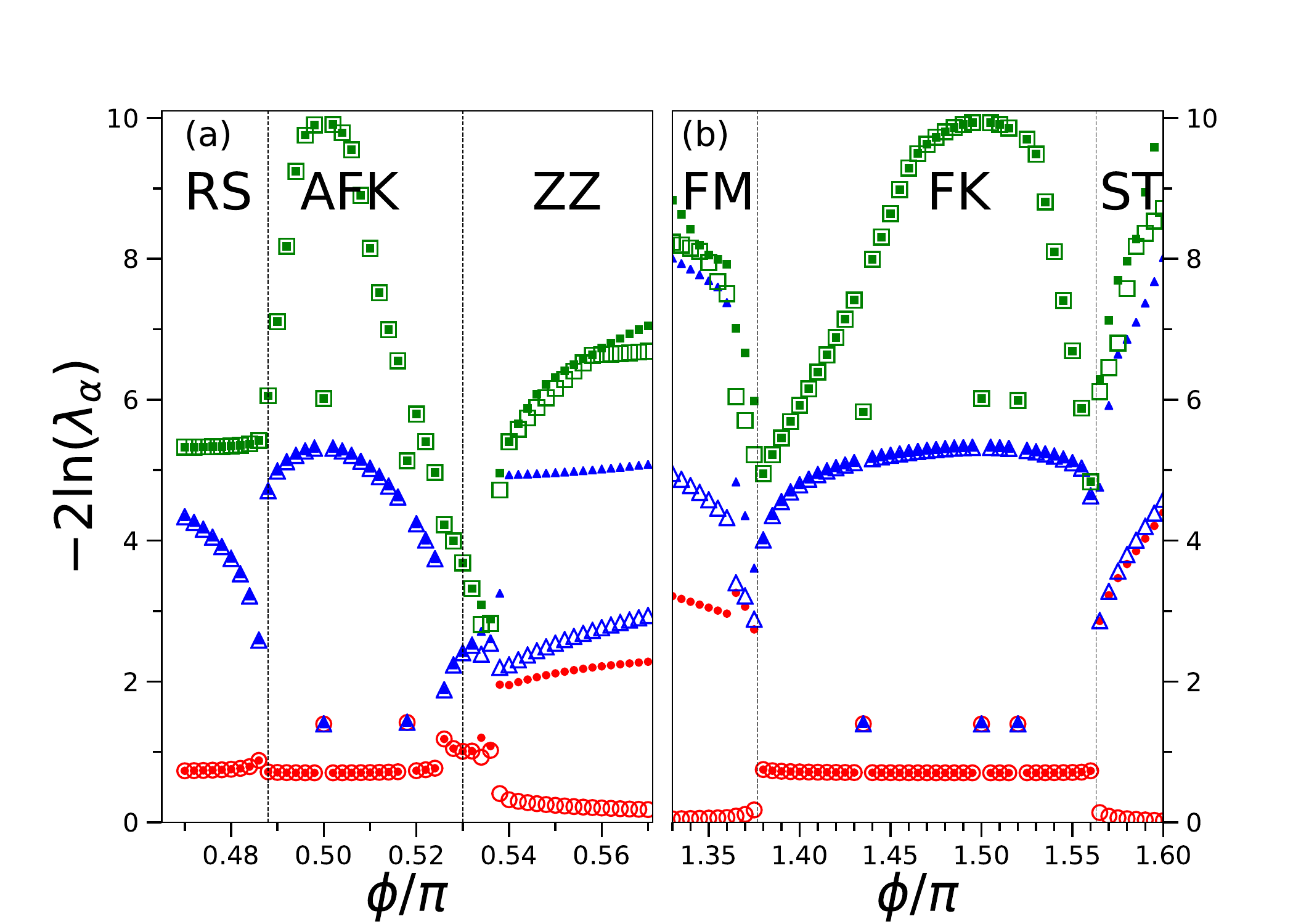}
\caption{\label{fig:ES} (Color online) Entanglement spectrum (ES) partitioned with
a cut of one middle rung is shown around 
(a) the AFK and (b) FK phases with $N=400$ and OBC.
 Open and filled red circles, open and filled blue triangles, and open and filled green squares correspond
to the 1st to 6th eigenvalues, respectively. See the main text for implications of the ES results. }
\end{figure}

Numerical results for $\langle {\cal O}^z\rangle$ as well as the gaps $\Delta E_1$ and $\Delta E_2$ to the first and second excited states
are shown near the AFK and FK phases in Fig.~\ref{fig:AFKLOzString}. 
Clearly, $\langle {\cal O}^z(r)\rangle$ attains the extremal values of $\pm 1$ in both cases exactly at the Kitaev points even though we have verified that 
all usual spin-spin correlators are {\it extremely} short-range.
Here, $\langle {\cal O}^z\rangle$ is obtained from the $\langle \tilde S^z_2\tilde S^z_{2+r}\rangle$ correlation function but we have checked that using
$\langle \tilde S^z_{N/2}\tilde S^z_{{N/2}+r}\rangle$ only makes minor changes. In the $N\to\infty$ limit the SOP remain finite in the FK and AFK phases
with $\Delta E_1$ and $\Delta E_2$ disappearing at the quantum critical points.

{\it Entanglement spectrum --} 
A topological phase is characterized by a double degeneracy of the entire entanglement spectrum (ES)~\cite{Li2008,Pollmann2009} obtained from the Schmidt coefficients $\lambda_\alpha$
of the partition.
Indeed the ES spectrum shown in Fig. ~\ref{fig:ES} is doubly degenerate for both AFK and FK, and RS  when partitioned with one middle-rung cut as shown in the 12-site ED with open boundary conditions (Fig. 9 in the SM).
  The middle-rung cut is important to generate the double degeneracy of
  the ES because a pairing term of Majorana fermions occurs via the dangling sites in $H_{d-Z}$ as shown in the SM.
  Thus, without the middle rung cut, the degeneracy is not expected.
  We have indeed confirmed that a vertical cut, not involving the middle-rung, does not give the ES degeneracy.

Note that in the FK and AFK the lowest ES eigenvalue is sporadically near four-fold degenerate. This tendency is more pronounced at smaller system sizes
and presumably arises from the near two-fold degeneracy of the ground-state~\cite{SM}.
The transition between the FK and ST/FM  as well as the AFK-ZZ transition is signalled by the disappearance of the double degeneracy present in the FK. However,
with our cut the RS-AFK transition is between two phases {\it both} with doubled ES that can only be differentiated by the SOP.
It is also important to differentiate the different natures of RS and AFK/FK. The edge states of RS are $S=\frac{1}{2}$, as 
they appear when a singlet formed on the middle rung is cut, while the edge states of AFK/FK are Majorana fermions. They are fractionalized excitations of $S=\frac{1}{2}$, similar to the original Kitaev honeycomb model.

The ES flow at the AFK-ZZ is rather unusual.
There is a disruption around 0.515$\pi$ due to a ground state level crossing.
Then at the AFK-ZZ transition
$\phi\sim 0.53\pi$ there is no noticable change in the degeneracy, instead, at the larger $\phi$ of 0.535$\pi$
the double degeneracy abruptly disappears.
This point shifts toward the actual transition point, as the system size increases
implying that the degeneracy splitting occurs only when the system size is larger than the correlation length. 
A further study is required to fully understand the nature of the AFK-ZZ transition.
        
{\it Summary and Discussion} --
Here we have proven the existence of a non-local SOP in the ladder of the Kitaev model which is
one of the unique quantities that characterize topological QPTs. 
We have shown that this SOP is non-zero in the AFK and FK phases of the KJ ladder both with and without Heisenberg term differentiating them from other nearby phases.
This is in contrast to the current understanding that these phases are disordered.\cite{Feng2007}
Furthermore, the transition boundaries surrounded by the AFK and FK phases are 
very similar to those of the 2D honeycomb limit\cite{ChaloupkaPRL2013} 
implying that the transitions are determined by the closing of the gap of the phases.

The ladder ZZ, ST, and RS phases will go through QPTs since a local order parameter associated with a magnetic order occurs in the 2D limit. 
On the other hand, AFK and FK phases become the gapless Kitaev spin liquid in the 2D limit.
While one expects a {\it topological} QPT, as the system
goes from a low dimensional gapped phase to a higher dimensional gapless non-Abelian phase, a long-range entanglement
seems to develop only in the true 2D limit.
Such development of long-range entanglement starting from the gapped phase of the ladder with its characteristic SOP to the 2D Kitaev spin liquid is a particularly interesting question for a future study.
Additionally, further studies on the Kitaev model including other interactions and/or magnetic field using the ladder geometry will advance
our understanding of Kitaev materials.

\begin{acknowledgments}
This research was supported by NSERC and CIFAR. Computations were performed in
part on the  GPC  and  Niagara  supercomputers  at  the  SciNet  HPC
Consortium. SciNet is funded by: the Canada Foundation for Innovation under the
auspices of Compute Canada; the Government of Ontario;  Ontario  Research  Fund
- Research  Excellence; and the University of Toronto.  Computations were also
performed in part by support provided by SHARCNET (www.sharcnet.ca) and
Compute/Calcul Canada (www.computecanada.ca). Part of the numerical
calculations were performed using the ITensor library (http://itensor.org)
typically with a precision of $10^{-11}$ and truncation errors not exceeding
$10^{-9}$ for the ground-state, $10^{-6}$ for excited states.
\end{acknowledgments}

\bibliography{references}

\end{document}